\documentstyle [fleqn,11pt] {article}
\begin{document}

\begin{titlepage}
\hspace{1cm}  \hfill ZU-TH-32/92
\vspace*{2cm}
\begin{center}{\huge MICROWAVE ANISOTROPIES FROM TEXTURE
	\vspace{.5cm}\\

	       SEEDED STRUCTURE FORMATION} \vspace{3.2cm}  \\

{\Large Ruth Durrer, }\\
Universit\"at Z\"urich, Institut f\"ur Theoretische Physik,
CH-8001 Z\"urich  Switzerland\\
{\Large Armando Howard}\\
Princeton University Observatory, Peyton Hall, Princeton,
NJ 08544 USA\\
 and\\
{\Large Zhi--Hong Zhou, }\\
Universit\"at Z\"urich, Institut f\"ur Theoretische Physik,
CH-8001 Z\"urich  Switzerland
\vspace{2.3cm} \end{center}
\begin{abstract}
The cosmic microwave anisotropies in a scenario of large scale
structure formation with cold dark matter and texture are
discussed and compared
with recent observational results of the COBE satellite.
A couple of important statistical
parameters are determined. The fluctuations
are slightly non gaussian. The  quadrupole anisotropy is
$1.5\pm 1.2\times 10^{-5}$ and the fluctuations on a angular scale
of 10 degrees are  $ (3.8\pm 2.6)\times 10^{-5}$.
The COBE are within about one standard deviation of the typical
texture + CDM model discussed in this paper. Furthermore, we calculate
 fluctuations on intermediate scales (about 2 degrees) with the
result $\Delta T/T(\theta \sim 2^o) = 3.9\pm 0.8)\times 10^{-5}$.

Collapsing textures are modeled by spherically symmetric field
configurations. This leads to uncertainties of about a factor of~2.
\end{abstract}
\end{titlepage}

\section{Introduction}
Global texture\cite{Tu}
is the latest of a set of models based on a simple
physical idea: the universe begins in a hot, homogeneous state
and then, as it cools,
undergoes a symmetry breaking phase transition
that leads to the formation of topological defects\cite{Ki}.  These
defects induce perturbations that seed the formation of
galaxy and large-scale structure.
Unlike stable defects (cosmic strings, domain walls and monopoles),
 textures are short-lived.
In a light-crossing time, they collapse
producing a singularity for one instant of time
and radiate their energy in a spray of Goldstone bosons.
They do, however, persist long enough to leave
an imprint as compensated perturbations in the matter and radiation
density.
The simplicity of the texture scenario is appealing, the model
has only a single adjustable parameter, the scale of symmetry
breaking, which we normalize to obtain the observed galaxy-galaxy
correlation function.

In a series of papers, we and our collaborators have
examined the texture-seeded $\Omega = 1$ CDM-dominated
cosmogony and found promising results.
Analytical studies of galaxy formation in this
scenario suggest that the model has several highly
attractive features \cite{GST}:
early star formation, which
could reionize the universe and smear out fluctuations in
the microwave background; early formation of galactic
spheroids could
account for their high mean density without significant
dissipation and would provide
a fertile environment for the formation of a massive black hole that
could power quasars;  a galaxy mass function whose slope and
amplitude is consistent with observations; and galaxy formation
at moderate $z$. Numerical simulations of the growth of large scale
structure in the texture scenario find striking consistency between
theoretical predictions and observations of large scale structure
\cite{Park,Cen,GST2}:
the galaxy-galaxy correlation function is consistent with
observation; the derived galaxy mass function is well-fit by
observations; the model produces coherent velocity fields on large
scales and the angular correlation  function of galaxies is
consistent with observations.

In this paper, we calculate the signature of microwave background
fluctuations induced by texture on the scales probed by COBE.
In section~2, we present our formalism for calculating the
response of dark matter, baryons and photons to the collapse
of a single spherically symmetric texture in an expanding
universe.  These calculations extend earlier work \cite{TS,Du}
that calculated the
fluctuations induced by a texture in flat space.  In section~3, we
describe how we sum over the contributions from many textures and
make a synthetic map of the microwave background.  Section~4
presents our numerical results and conclusions.

\section{The Effects of the Collapse of a Texture}

Texture arises when the symmetry breaking ``Higgs" field chooses
to lie at different places on its degenerate vacuum manifold in
different regions of space during the phase transition.
This can produce configurations in which the Higgs field has
non-trivial
winding number on the vacuum manifold, which we call ``knot".
Such a knot is unstable to collapse --- it shrinks until the
gradient of the
Higgs field becomes large enough that the field lifts off  the
vacuum manifold for an instant, undoing the knot.
This happens when the knot shrinks to a microscopic size, of order
the inverse symmetry breaking scale.
After the knot unwinds, the field is  free to radiate away
completely.

The cosmological evolution of texture is remarkably simple: in
any given horizon volume, there is a fixed probability of the field
configuration being wound into a knot\cite{Tu,STPR}.
Once there is sufficient winding number density within the horizon,
the knots collapse at the speed of light. As the horizon size grows
 and
``connects" new regions, the process repeats in a self similar
manner,
with new knots continually collapsing on progressively larger
scales. As they collapse, unwind, and radiate away, the knots
generate a gravitational field that perturbs the surrounding matter.

In this paper we study the simplest case of texture: that formed
by a completely broken global $SU(2)$ symmetry.
It is important to notice, however, that the general picture of
cosmological texture evolution just discussed is expected to be
the same regardless of the ``species" of texture.
Thus we expect many of the main features of the resulting structure
 formation to be species independent.
Furthermore, it turns out that for any texture the dynamics of the
field is dependent only on the geometry of the vacuum manifold,
and cannot be affected by the shape of the Higgs potential or
even the symmetry breaking scale.
The latter does enter our scenario as the one ``tunable" parameter,
but affects only the {\it amplitude} of the induced fluctuations,
not the pattern.
We will see that normalizing the fluctuations such that the
resulting
cosmic structure matches well with observation implies a symmetry
breaking scale quite natural for GUT textures.

\subsection{Formalism}

In this section, we will present a formalism for calculating the
response of matter and radiation to the collapse of a single texture
knot.    Throughout this paper, we will assume a flat Friedman
universe,
$\Omega_{tot} =1$, dominated by CDM and  we work with conformal time
so that the background metric is given by
\[ ds^2 = a^2(-dt^2 +d\mbox{\boldmath $x$}^2)  \; .\]

We will model the effects of texture by considering the response
of matter to the collapse of a single spherical texture.  This
approach has proven fruitful in describing the response of dark
matter to texture\cite{GST2}.
The spherically symmetric ansatz for an ${\bf S}^3$ texture
(i.e. a   configuration of a scalar field living on ${\bf S}^3$
which winds once around ${\bf S}^3$ as $r$ goes from $0$ to
$\infty$)  unwinding at a given time $t=t_c$ is \cite{TS}
\begin{equation} \mbox{\boldmath $\phi$} =
\eta^2(\sin\chi\sin\theta\cos\varphi,
	\sin\chi\sin\theta\sin\varphi,
	\sin\chi\cos\theta, \cos\chi)  \label{2phi} \; , \end{equation}
where $\theta \; , \; \varphi$ are the usual polar angles and
$\chi(r,t)$ has
the following properties
\[ \begin{array}{ll}
     \chi(r=0,t<t_c) & = 0 \\
     \chi(r=0,t>t_c) & = \pi \\
     \chi(r=\infty,t) & = \pi \; .
  \end{array}  \]
Since the texture seed is already a perturbation, it can be evolved
according to the background equation of motion,
$\Box\mbox{\boldmath $\phi$}-
 (\mbox{\boldmath $\phi\cdot$}\Box\mbox{\boldmath $\phi$})
 \mbox{\boldmath $\phi$} =0$,
which yields
\begin{equation} \partial_t^2\chi + 2(\dot{a}/a)\partial_t\chi -
\partial_r^2\chi -{2\over r}
    \partial_r\chi  = -{\sin 2\chi \over r^2} \; . \label{2chi} \end{equation}
The energy momentum tensor of the texture field,
$ T_{\mu\nu} = \partial_{\mu}\mbox{\boldmath $\phi$}\cdot
 \partial_{\nu}\mbox{\boldmath $\phi$} -
  {1\over 2}g_{\mu\nu}\partial_{\lambda}\mbox{\boldmath$\phi$}\cdot
 \partial^{\lambda}\mbox{\boldmath $\phi$}$,
can be parametrized in terms of dimensionless functions,
\[ T_{00} = 4\eta^2f_{\rho}/l^2 \; , \;\;\;
   T_{0i} = -4\eta^2 f_{v,i}/l \; , \]
\[ T_{ij} = 4\eta^2[ (f_p/l^2 - {1\over 3}\triangle f_{\pi})\delta_{ij}
		+ f_{\pi},_{ij}]  \; , \]
where (\begin{equation} f_{\rho}/l^2 = 1/8[(\partial_t\chi)^2
+(\partial_r\chi)^2 +
	{2\sin^2\chi\over r^2}]
       \label{2fro} \; , \end{equation}
\begin{equation} f_p/l^2 = 1/8[(\partial_t\chi)^2 -{1\over 3}(\partial_r\chi)^2
      - {2\sin^2\chi\over 3r^2}]  \label{2fp} \; , \end{equation}
\begin{equation} f_v/l  = {1\over 4}\int_r^\infty
(\partial_t\chi)(\partial_r\chi)dr \label{2fv} \; , \end{equation}
\begin{equation} \triangle f_{\pi} = {1 \over 4}[(\partial_r\chi)^2
-{\sin^2\chi\over r^2}
	] +  {3\over 4}\int_{\infty}^r[(\partial_r\chi)^2 -
       {\sin^2\chi \over r^2}]{dr \over r}  \label{2fpi} \; . \end{equation}
The length parameter $l$ is introduced to keep the functions
$f_{\rho}$ to
$f_{\pi}$ dimensionless. A convenient choice is to set $l$ equal the
typical size of the perturbations, e.g., the horizonsize at texture
 collapse.
$f_{\rho}/l^2$ and $f_p/l^2$ are the energy density and isotropic
pressure
of the texture field.  $f_v$ is the potential of the velocity field
and $f_\pi$ is a potential for  anisotropic stresses.

Cosmological perturbations are usually split into scalar, vector and
tensor contributions which do not couple to first order (here the
terms
scalar vector and tensor refer to the transformation properties
 on  hypersurfaces of constant time).
The split into scalar, vector and tensor perturbations is nonlocal
and thus  acausal. But spherically symmetric perturbations are of
course always of scalar type (due to the adoption of spherical
symmetry,
we loose, e.g., all information about gravity waves produced during
 the collapse.). In this case it is  very convenient to apply
gauge--invariant cosmological perturbation theory which is adapted
to a split into scalar, vector and tensor modes  \cite{Ba,KS,DS,Du}.
In this paper we use the formalism and notation introduced in
\cite{Du}.

 The gravitational field  can be described in terms
of the gauge--invariant Bardeen potentials $\Phi$ and $\Psi$.
The explicit
definition of the Bardeen potentials in terms of the metric
perturbations
is given in \cite{Ba,KS,DS,d92}. For perturbations which are
much smaller than the size of the
horizon, $l\ll l_H$, $\Psi$ corresponds to the Newtonian potential,
 and
$\Phi$ is related to the perturbation of the 3--curvature on
hypersurfaces of constant time:
\[ -(4/a^2)\triangle\Phi = \delta R^{(3)} + {\cal O}(l/l_H) \; . \]
Hence for subhorizon perturbations  of ordinary  matter
($|T^{00}| \gg |T^{ij}|$), the well known Newtonian limit of
Einstein's equations (see, e.g, \cite{St}) yields
\[ \triangle\Phi = -\triangle\Psi \; . \]

Einstein's equations determine the Bardeen potentials induced
by the matter perturbations \cite{Du}. In addition to texture we
want to describe dark matter and radiation. Since dark matter
has zero
pressure, we just need a variable describing its density
perturbation.

By $\rho_d$ we denote  the density of  dark matter. $D$ is
a gauge invariant combination of the density perturbation,
$\delta = (\rho_d-\rho_{d0})/\rho_{d0}$, the potential
for the peculiar velocity,
$lv~,~~ v_i = -lv,_i$ and geometric terms.

In this variable,
the evolution of  dark matter fluctuations is governed by
\begin{equation} \ddot{D} + (\dot{a}/a)\dot{D} - 4\pi G\rho_da^2D =
       \epsilon(f_{\rho} +3f_p)/l^2 \; , \label{2D} \end{equation}
where $\epsilon=16\pi G\eta^2$. A derivation of this equation is
presented  in\cite{Du}.
After decoupling, the  evolution of radiation
 is given by the one particle Liouville equation. Its first order
perturbation yields a differential equation for $\mbox{$\cal M$}$, which is a
gauge--invariant version of the perturbation of the energy
integrated photon distribution function.
\begin{equation} \mbox{$\cal M$} = {4\pi\over \rho }\int_0^\infty {\cal
F}p^3dp  ~.
	\label{mm}\end{equation}
$\mbox{$\cal M$}$ evolves in response to
fluctuations in both the $\Phi$ and $\Psi$ potentials:
\cite{Du89,d92}
\begin{equation} \dot{\cal M} +\gamma^i\partial_i\mbox{$\cal M$} =
4\gamma^i\partial_i(\Phi-\Psi)~.
	\label{2li}  \end{equation}
The unit vector {\boldmath $\gamma$} denotes the direction cosines of the
photon
momentum.  In spherically symmetric systems, the photon evolution
equation (\ref{2li}) reduces to,
\begin{equation} \dot{\mbox{$\cal M$}} +\mu\partial_r\mbox{$\cal M$} +
{1-\mu^2\over r}\partial_{\mu}\mbox{$\cal M$} =
 4\mu\partial_r(\Phi-\Psi) \label{2mmc} \end{equation}
where $\mu$ is the direction cosine of the photon momentum in
the direction of {\bf r}.

The photon evolution equation is more transparent in
characteristic coordinates, $(t,\tau,b)$ where
$b = r\sqrt{1-\mu^2}$ is the impact parameter and $\tau=t-r\mu$
is the time at which the photon is at its minimum distance from
the texture, the 'impact time'.
In these variables (\ref{2mmc}) simplifies to
\begin{equation} \partial_t\mbox{$\cal M$}(t,\tau,b) =  4\mu\partial_r(\Phi
-\Psi) ~, \end{equation}
so, that
\begin{equation}  \mbox{$\cal M$}(t,\tau,b) =
	\mbox{$\cal M$}(t_o,\tau,b) + 4 \mu \int_{t_o}^t dt'\partial_r(\Phi
	-\Psi)~.
     \label{2lis}  \end{equation}

To write down the perturbed Einstein equations we have in principle
to calculate the energy momentum tensor of radiation from \mbox{$\cal M$}. But
since we are only interested in late times, $\rho_d\gg\rho_\gamma$,
we may neglect the contribution of radiation to the density
perturbation. The potential $\Phi$ is then determined by the
texture and dark matter perturbations alone \cite{Du}:
\begin{equation} \triangle\Phi = -\epsilon(f_{\rho}/l^2+ 3(\dot{a}/a)f_v/l)
   -4\pi Ga^2\rho_dD  \;  ,\label{2Phi0}\end{equation}

On the other hand, since dark matter does not give rise to
anisotropic stresses, we have to take into account the contribution
of radiation to the latter.
To calculate the anisotropic  stresses of radiation we recall
the definition of its amplitude, $\Pi$
(see e.g. \cite{Du89,d92}):
\[ \delta T_i^j - {1 \over 3}\delta T_l^l\delta_i^j = p[\Pi_{,i}^{,j} -
     {1\over 3}\triangle\Pi\delta_i^j] ~.  \]
The anisotropic contributions to the energy momentum perturbations
of  photons are given by
\[ \delta T_i^j - {1 \over 3}\delta T_l^l\delta_i^j = {\rho_\gamma\over 4\pi}
  \int(\gamma_i\gamma^j-{1\over 3}\delta_i^j)\mbox{$\cal M$} d\Omega ~.\]
Using these equations and spherical symmetry one finds
\begin{equation} \Pi''-\Pi'/r = {9\over 4}\int_{-1}^1(\mu^2-{1\over
3})\mbox{$\cal M$} d\mu
  =: {9\over 2}M_2(r,t)  ~.  \label{2Pi}  \end{equation}
This anisotropy and the anisotropy of the texture field contribute
to the sum of the two Bardeen potentials \cite{Du}

\begin{equation}  \triangle(\Psi +\Phi) = -2 \triangle(\epsilon f_\pi +{4\over
3}\pi Ga^2
	\rho_\gamma\Pi)    ~. \label{2Psi}   \end{equation}

Inserting the results for $\partial_r(\Phi-\Psi)(r(\tau,\mu),t)$ at a
given
time step, we can evolve the brightness function $\mbox{$\cal M$}$ according to
(\ref{2lis}). The function $\mbox{$\cal M$}$ is very simply related  to the
temperature
perturbation in a given direction in the sky: To see this, let
us express the perturbation of the photon distribution function as
a perturbation of the temperature:
$f = f_{(0)} +\delta f = f_{(0)}(p/T(t,r,\gamma))$ so that $\delta f =
	{df_{(0)}\over  dv}\delta T/T$. If we know neglect the
difference between gauge--invariant
and gauge dependent variables and set $\delta f \equiv {\cal F}$,
we find from (\ref{mm}) the simple relation $\mbox{$\cal M$} = 4\delta T/T$. In
the gauge--invariant framework, this just means that $\mbox{$\cal M$}/4$ is a
gauge--invariant variable describing the perturbation of the
temperature in a given direction (for
a more explicit discussion see \cite{d92}).

We would like to  choose initial conditions that are physically
plausible.  If the phase transition that produced texture occurred
in an initially uniform universe,  causality requires that there
are no geometry fluctuations well outside the horizon \cite{VS}.
This implies that initially, at $t_i\ll t_{col}$, we must require
$\Psi=\Phi = 0$.
We want to compensate as much as possible of the initial texture
fluctuations with an initial dark matter perturbation.
Hence, we use as initial conditions for the density field,
\begin{equation} D(r,t=t_i) = -{\epsilon \over 4\pi Ga^2\rho_d} (f_{\rho}/l^2+
	3(\dot{a}/a)f_v/l)
     ~.   \label{2D0}\end{equation}
This initial condition implies that metric fluctuations are induced
by the differences between the texture equation of state and
the equation of state of the background matter.
This choice yields $\Phi = 0$ at $t=t_i$, but  not  $\Psi = 0$.
Due to its equation of state the dark
matter cannot compensate the anisotropic stresses of the texture.
We thus must compensate them by an initial photon perturbation.
For $\Psi$ to vanish,  we have to require according to (\ref{2Psi})
\[ \Pi = -{3\epsilon\over 4\pi Ga^2\rho_\gamma}f_\pi~.\]
Clearly this does not lead to a unique initial condition for \mbox{$\cal M$},
but if we in addition require the zeroth and first moments of
\mbox{$\cal M$}~ to vanish it is reasonable to set
\begin{equation} \mbox{$\cal M$}(r,\mu,t=t_i) = -{15\epsilon\over 8\pi
Ga^2\rho_\gamma}
	(\mu^2-{1\over 3})
   (f''_\pi -{1\over r}f'_\pi)
  ~. \label{2mm0} \end{equation}
Together with initial conditions for the texture field
$\chi(r,t=t_i)$,
the requirements (\ref{2D0}) and (\ref{2mm0}) and the evolution
equations (\ref{2chi}) to (\ref{2Psi}) determine the system which
we solved numerically.

\subsection{Texture Collapse in Flat Space}

Before we present numerical methods and  results, let us briefly
discuss the flat space case, where, if we ignore the response of
  dark matter to the texture, the microwave fluctuations can be
calculated analytically.
The Bardeen potentials of a  self similar texture collapsing at
$t_c=0$ are  \cite{Du}
\[ \Phi = -{\epsilon \over 4}\ln({r^2+t^2\over l^2})~~,~~~~~
    \Psi = {\epsilon \over 4}\ln({r^2+t^2\over t^2})~~, \]
so that (\ref{2mmc}) turns into
\begin{equation} \partial_t\mbox{$\cal M$} = {-4\epsilon(t-\tau)\over
b^2+(t-\tau)^2 +t^2} =
	g(t,\tau,b)~.\end{equation}
Integrating that over time one finds
\begin{eqnarray}  \mbox{$\cal M$}(t,\tau,b) &=& \mbox{$\cal M$}(t_i,\tau,b) +
{2\epsilon}
[{\tau\over\sqrt{\tau^2+2b^2}}
    (\arctan{2t-\tau\over\sqrt{\tau^2+2b^2}} -
  \arctan{2t_i-\tau\over\sqrt{\tau^2+2b^2}})  \nonumber \\  &&
- \log({(2t-\tau)^2 +\tau^2+2b^2
     \over(2t_i-\tau)^2 +\tau^2+2b^2})]  \label{3mf} \end{eqnarray}a
We are interested in the change of the distribution function due
to the collapsing texture and thus want to perform the limit
$t\rightarrow\infty~,~t_i\rightarrow
 -\infty$, with $\lim_{t_i\rightarrow-\infty}\mbox{$\cal M$}(t_i,\tau,b) =0 $.
The log term in (\ref{3mf}) then contributes a constant
$C= \lim(\log (t^2/t_i^2))$ which crucially depends upon how we
perform this limit. It can take any value
$-\infty \le C\le \infty$.
This can happen because the flat space texture is an unphysical
infinite energy solution. If we would change the energy
momentum tensor  in a way that the texture would be 'born' some
time in the finite past, or if we would compensate it in a
consistent way as we
do it in the expanding universe, this problem would disappear. For
simplicity, and since we are not interested in a $\tau$ and $b$
independent constant, we just perform the limit $t=-t_i\rightarrow\infty$
which yields $C=0$. Then the final change of the distribution
function is
\begin{equation} \Delta\mbox{$\cal M$}(\tau,b) =
2\epsilon\pi{\tau\over\sqrt{\tau^2+2b^2}}  ~.\end{equation}
Since $\mbox{$\cal M$}$ describes the fluctuations in the energy density,
$\Delta T/T = \Delta\mbox{$\cal M$}/4$.

In a flat universe the signal from a texture collapsing at $t=0$
as seen from an observer at time $t_o$ and distance $r_o$ would
thus be
\begin{equation} {\Delta T\over T}(\theta) = \epsilon\pi/2{t_o
-r_o\cos\theta\over \sqrt{
     (t_o-r_o\cos\theta)^2 +2r_o\sin^2\theta}}  \end{equation}
where $\theta$ is the angle between the direction of the texture and
the line of sight. A similar calculation using a specific gauge but
performing a somewhat unphysical limit is presented in \cite{DHJS}.
Unlike in the expanding universe  (see 2.3), there
is no horizon present in these calculations, thus, photons that
pass the
texture long before or after collapse, $|t_o|\gg r_o$ are still
influenced by it and yield even a maximum temperature shift,
\[ {\Delta T\over T} = \pm  \epsilon\pi/2 ~.\]
In the expanding universe we expect $\Delta T/T$ to achieve a maximum
for  $r_o\approx t_o-t_c$ and to vanish for $t_o-t_c \gg r_o$.

\subsection{Texture Collapse in the Expanding Universe}

The coupled set of equations that describe the evolution of
texture, the fluctuations in the dark matter, the fluctuations
in the metric, and the fluctuations in the microwave
background in the expanding universe must be integrated numerically.
We use a second-order accurate leapfrog scheme to advance these
equations.

We will assume a specific set  of initial conditions for the
texture field.   One of the sources of uncertainty in our
calculations
of the microwave background is the choice of initial conditions.
We have found that different choices of initial conditions can
yield microwave fluctuations that differ by $\sim 50\%$ from
the values reported here.
We choose initial values for $\chi$ that confine the fluctuations in
$\chi$ to a region comparable to the horizon size at collapse:
\begin{equation} \chi =  A r/r_0 +  B (r/r_0)^2 + C (r/r_0)^3
\quad r < \alpha r_0  \end{equation}
$$ \chi = \pi \qquad \qquad r > \alpha r_0$$
where $A$, $B$ and $C$  are determined by the requirement
that $\chi-\pi$ and its first and second derivatives vanish at
$r=\alpha r_0$.
For $\alpha = 4/3$ this texture  collapses at $t \approx r_0$.
In contrary to the self similar flat space solution given in
\cite{TS,Du},  for this set of initial conditions,
$f_\pi$ does not vanish initially   and the
$\Psi(t_i)$ contribution must be compensated by a photon
anisotropy $\Pi$ as explained above.

Figure 1 shows the microwave background fluctuations induced by
the collapse of the texture as a function of $\tau$ for small impact
parameter for both the flat space solution (dotted line) and
the texture in the expanding universe.  The dotted line
in the figure is the flat space result from section 2.2.  Figure 2
shows the microwave background fluctuations induced by texture
collapse as a function of impact parameter.  Note that the
temperature fluctuations
are induced only for photons that pass within the event horizon of
the texture.

\subsection{Compensation}
In addition to the choice of sensible initial condition, we must
keep in mind that spherical symmetry in an acausal concept and
induces accelerations on super horizon scales. This is pictured
by the dashed line of Fig.~3.
There $\Delta T/T$ is shown as a function of the impact time
$\tau$ at constant time $t\approx t_c$ and impact parameter
$b\approx t_c/50.$.  One sees that photons with impact time
$\tau\approx -40.= -t_c$, i.e. photons which had a distance
 $r>t_c$ from the texture at the time of the big bang, $t=0$ and
always moved in the direction opposite to the texture
($\mu\approx 1$), still
experience nearly  half of the maximum energy shift even though
they have never been in causal contact with the region of texture
collapse (which at the
collapse time, $t=t_c$ encompasses a radius of $t_c$).

These acausalities would lead to an overestimate of $\Delta T/T$.
To prevent this we have to compensate in such a way that there
are no
accelerations on scales larger than the horizon. At each time we
thus set the
compensation radius equal to the size of the horizon,
$r_{cp} = l_H=t$.
We then set $f_\rho(r) = f_p(r) = f_v(r) = f_\pi(r) =0$ for
$r>r_{cp}$.
On small scales we subtract the mean of $f_\rho$ and $f_p$, so that
\[ \int_0^{r_{cp}}f_\rho =   \int_0^{r_{cp}}f_p =  0~.\]
A texture compensated in this way does not produce any
gravitational fields  on scales larger than $r_{cp}$.

We have added this compensation to the calculation of the texture
sources. The signal of a collapsing texture has then reduced to
the solid line in Fig.~3. Here photons which have never been and
will never be in causal contact with the texture ($\tau<0$) and
photons which are not yet in
causal contact with the texture ($\tau>2t_c=80$) do not acquire
any appreciable energy shift. A result which is much more
physical than the dashed line.

Using this compensation scheme, we assume that spherical symmetry
is a good approximation inside the horizon and that the stocastic
texture field outside the horizon can be neglected. The question
if this assumption is justified can only be answered by a full 3d
simulation like \cite{BR}.
We believe that the issue of compensation and the arbitrariness in
the initial conditions yield a factor of about 2  uncertainty in
our results (which is much bigger than our numerical errors). It
might well be, that by neglecting super horizon fluctuations
completely, we slightly underestimate
$\Delta T/T$ but most probably by less than a factor 2.

\section{Textures and the Microwave Sky}

In the previous section, we described how the collapse of a single
texture produces fluctuations in the photon temperature.  In this
section,
we sum the contribution of multiple textures and describe how to
construct microwave maps of the night sky.

Since COBE observations cover the entire celestial sphere, we
construct a numerical grid consisting of 1 square degree patches.
  These patches are arranged so that they cover equal areas and
each patch has roughly the same shape.

Since the texture fluctuations are in the linear regime, we assume
that the contributions of each texture to the fluctuations at
each point in the sky can be added independently.  We randomly throw
down textures everywhere within the event horizon using the
texture density distribution function \cite{STPR},
\begin{equation}{dn \over dt} = {1 \over 25} {1 \over t^4}, \end{equation}
follow the collapse of each texture and sum their contributions.
We include only contribution of textures that collapse after
recombination.
Textures that collapse earlier do not contribute significantly to
microwave fluctuations on scales accessible to COBE.

In order to simulate the COBE observations, the
map of the night sky  is smoothed with a Gaussian beam with a
FWHM of $7^\circ$,
the angular resolution of the DMR detector on COBE\cite{Sm,aps}.
After computing and removing the intrinsic dipole contribution
($10^{-5} - 10^{-4}$) and any monopole fluctuations, we then
compute the microwave quadrupole, the r.m.s. pixel-pixel
fluctuations and other statistics of the microwave sky.

The simulated 'COBE--map' is shown in Fig~4.

\section{Results}
The amplitude of the microwave background fluctuations depends
upon the scale of symmetry breaking associated with the texture.
If the texture model is  normalized so that it can reproduce
the galaxy-galaxy correlation function, then
	$\epsilon = 5.6 \times 10^{-4} b^{-1}$,
where $b$, the bias factor, is the ratio of the mass-mass
correlation
function to the galaxy-galaxy correlation function \cite{GST,PST}.
Comparison of the predictions of the texture model with
observations of clusters  \cite{Bart}
suggest that $b \approx 2-4$.
Hydrodynamical simulations of texture-seeded galaxy formation
\cite{Cen} suggest that $b \approx 2$.

Figure 5  shows the distribution of quadrupole values calculated from
  100 realizations of model.    Averaging over the
realizations, we find an r.m.s. value
for $Q$ of
\[ Q= (1.5\pm 1.2)\times 10^{-5}\epsilon_0~, \] where we have set
$\epsilon_0=\epsilon/2.8\times10^{-4}$
so that a bias factor of 2 and the above normalization  would
yield $\epsilon_0=1$. In 13 percent of the realizations the calculated
value of the quadrupole is below the COBE result
($Q=5\times 10^{-6}$). Since only a handful of texture knots
are the source of most of the large scale fluctuations, the
quadrupole
varies significantly from realization to realization.  The 95\%
confidence range is $3.5 \times 10^{-6} \epsilon_0$ to
$ 4\times 10^{-5} \epsilon_0$.
The COBE value is well within one standard deviation.

The texture-induced microwave sky shows non-Gaussian features
that distinguish it from the standard inflationary models.
Figure 6 shows the distribution of pixel-to-pixel fluctuations
calculated in one of the simulations.  The dotted line shows a
Gaussian whose dispersion matches the temperature dispersion of
this simulation.
The average temperature fluctuation averaged over 100 simulations is
\[(\Delta T/T)_{rms}(\theta=10^o) =  (3.8\pm 2.6)\times 10^{-5}\epsilon_0~. \]
This number is
consistent with estimate based on three-dimensional simulations of
texture induced microwave fluctuations\cite{PST,BR}.
Figure~6 shows that
the distribution of temperature fluctuations are only mildly non-
Gaussian, the skewness of the distributions averaged over 100
realizations
 is $-4\pm 2.3$ and
the kurtosis of the distributions is $32\pm 29$.

Considering these results, the recently reported results from the
COBE differential microwave background
radiometers [DMR] \cite{aps} are consistent with the texture +
CDM scenario.
The experiment has measured a value of
$Q = (0.5\pm 0.1) \times 10^{-5} $
for the microwave quadrupole and $(\Delta T/T)_{rms} = (1.1\pm 0.18)
 \times 10^{-5}$ for the pixel to pixel fluctuations in the
microwave maps.
Both these values lie within one standard deviations from the
results of
more than 100 realizations of the texture scenario.
(Recent reexaminations of the COBE results suggest that
the quadrupole may be somewhat lower \cite{Gould}.)
For the texture scenario, this implies that
$\epsilon \approx 3 \times 10^{-4}$ and a bias factor of
$b\approx 2$ are
compatible with both, the large scale structure formation and
the microwave background anisotropies produced in this scenario.

Our simulation does not take into account the baryon/photon coupling
before recombination and does not properly model the process of
recombination itself. It is thus not  suited to account for textures
which collapse before recombination, i.e., for angular scales beyond
about $\theta_{min}=2^o$. At this minimal resolution, the fluctuations
in  our simulations are

\[(\Delta T/T)_{rms}(\theta=2^o) =  (3.9\pm 0.8)\times 10^{-5}\epsilon_0~. \]
It is interesting that the cosmic variance (i.e. difference in values
over our 100 realizations for this angle now has really dropped. This is
clear, since more than 1000 textures have contributed to this signal
in each map. Our number does contradict the south pole result \cite{Gai}
but is in marginal agreement with  the recent MAX experimant (third flight)
\cite{MAX}. In Fig~7 the autocorrelation function is shown with $1\sigma$
upper and lower limits. On small scales, it is typically a factor
2 -- 3 higher than the COBE result \cite{aps}. This shows that the texture
scenario (without reionization) marginally conflicts observations on small
scales ($\theta\le 5^o$).

In Fig.~8 the amplitudes of the spherical
harmonics are presented. To specity our conventions, we define them here:
\begin{equation} (\Delta T/ T)(\vartheta,\phi)=
 \sum_{l=0}^\infty\sum_{m=-l}^la_{lm}Y_{lm}(\vartheta\phi) , ~~~\mbox{
 where}\end{equation}
\[ a_{lm} = \int d\Omega (\Delta T/T)Y^*_{lm} ~. \]
The amplitudes $C_l$ are then given by
\begin{equation} C_l = {1\over 2l+2}\sum_{m=-l}^l|a_{lm}|^2 ~. \end{equation}
In Fig~8 we present in addition to the values $<C_l>$ averaged over 100
realizations also their $1\sigma$ upper and lower limits. Furthermore, the
analytical fits for power law spectra with $n=1.5$ (short dashed) and
$n=1$ (long dashed) are plotted. A simple investigation leads to a $1\sigma$
limit of the spectral index of about $n=1.2\pm 0.3$, although our numerical
result represents definitely not a pure power law.

Up to now we have only considered the statistical variations of our
results which vary accidentally from realization to realization, i.e.,
cosmic variance. It is however important to
note that due to the uncertainties
in the initial configurations of the textures and the approximations
inherent in modelling the texture as spherically symmetric,  these
estimates of the microwave background fluctuations are uncertain
by a factor $\sim 2$, representing a systematic error.
Since in this approximation we completely neglect
 all non topological fluctuations of the scalar field, we probably
underestimate the induced fluctuations. On the other hand, a
different
choice for the scalar field leading to a different vacuum manifold
$M$ may as well alter the results by a factor of 2 (by altering the
probability of texture formation). In addition, we throw down the
textures randomly, uncorrelated whereas 3d numerical simulations hint
that they are probably anticorrelated.

Because of the non-Gaussian nature of the initial conditions,
galaxy formation begins much earlier in the texture model than
in Gaussian theories of structure formation.  By $z= 50$, a fraction
$6 \times 10^{-4}$
of the mass of the universe has formed non-linear objects
of mass greater than $10^6$M$_\odot$ --- these objects
may have reionized the universe.  If 10\% of the baryons
in these objects are ``burnt" in high mass stars, then
the resultant energy, 60 eV per baryon
 is sufficient to
reionize the universe.  The optical depth of an ionized universe is
\begin{equation} \tau(z) = 0.9\cdot 10^{-3}({h_{50}\Omega_b\over 0.05})
	[(1+z)^{3/2}-1] \end{equation}
so that we get an optical depth of 0.33
back to $z=50$  in an $\Omega_b = 0.05$, $h_{50}=1$, and
$\Omega = 1$ universe. A very  rough estimate of the effects of
reionization by smoothing each texture with a smoothing scale
of about the horizon size at $z_{dec}$,
 where $z_{dec}$ is determined
by $\tau(z_{dec}) = 1$, thus
\[ z_{dec} = 107\left({0.05\over h_{50}\Omega_b}\right)^{2/3} ~. \]
This corresponds to an angular scale of
\begin{equation}  \theta_{smooth} = t(z_{dec})/t_0 = (1+z)^{-1/2} \approx 5.7^o
\left({\Omega_b h_{50} \over 0.05}\right)^{1/3} ~.\end{equation}
This smoothing length is somewhat larger than the resolution scale
of COBE. If the formation of objects leads to reionization prior
to $z_{dec}$ this would suppress microwave background fluctuations
on the smallest scales probed by COBE ($\sim 3^o$), but would
not effect the fluctuations on the larger scales ($\ge 10^o$)
discussed above.

We  conclude again that our simulations indicate that the texture
scenario can reproduce the CBR anisotropies measured by COBE and no
 pronounced inconsistencies with the COBE data are found.

Due to the fact that a collapsing texture is a rare event, the
standard deviations of fluctuations increase strongly towards larger  scales.
This leads to an increase by nearly a factor of 2 from the $1-\sigma$ lower
limit of the fluctuation amplitude at 10 degrees to the corresponding limit
at 2 degrees. This can be compensated by the damping
of fluctuations due to reionization which can also affect scales
which are
somewhat larger than $\theta_{smooth}$.

To further test this scenario it is important, in addition to a full
three dimensional investigation of the CBR anisotropies, to discuss
the damping by reionization in  detail for
anisotropies on intermediate scales, $1^o$ to $5^o$.

Our work confirms that the texture scenario is a viable candidate for
large scale structure formation which cannot be excluded by present
observations. Its CBR anisotropies can be distinguished from those of
a scenario with initial fluctuations induced due to an inflationary
 phase by two statistical criteria: The fluctuations are slightly
non Gaussian and
they increase somewhat towards smaller angular scales.
A speciality of the texture scenario is furthermore the very distinct
signal which one singular texture collapse would imprint on the
microwave
sky. Depending on the time of texture collapse and on our distance
to it, we would see a roughly spherical ring of negative, positive or
negative--positive fluctuation, see Fig.~9. Given the statistical
distribution of textures (23), we expect to observe about $5-10$
such rings  with an angular diameter around $10^o$.

It is clear that although as a function of the impact parameter the phton
signal shows a blueshift followed by a redshift, what is observed at a
given moment in the sky is either dominated by redshifted or by bueshifted
photons, since most photons arriving at the observer today have impact times
which differ by much less than the typical size of the texture. Due to
spherical symmetry, the signals form rings in the sky. In a more realistic
 3d simulation these rings would be deformed. The
blueshifted rings are more affected by compensation and thus less
pronounced as seen the map (Fig.~4).
 \vspace{1cm}\\
{\large\bf  Acknowledgement}\\
We would like to thank Lyman Page, Norbert Straumann and Neil Turok
for stimulating discussions. We are very grateful to Ken Ganga who
helped us producing the color map and David Spergel who participated in
the first part of this project.\\
R.D. and Z.--H.Z. acknowledge support from the Swiss NSF.

\newpage

\noindent {\Large \bf APPENDIX}
\vspace{1cm}

\appendix
\section{Damping by photon diffusion}
In this appendix we want to estimate the damping of CBR
fluctuations in an ionized plasma.
For simplicity we work here in physical, not conformal time and
we always consider a flat, matter dominated universe, so that
$  t = t_0/(1+z)^{3/2} $ and $\rho_\gamma <\rho_b$.

For a perturbation with wave number $k$ which is significantly
smaller than the mean free path, the damping rate is then
\cite{Pee}
\begin{equation} \gamma \approx k^2t_T/6 ~,\end{equation}
where $t_T = 1/(\sigma_Tn_e)$ is the mean free path of the photons.
We now consider a texture which collapses at time
	$t_c = t_0/(1+z_c)^{3/2}$.
The total damping, $\exp(-f)$, which this perturbation experiences is
given by the integral
\begin{equation} f \approx \int_{t_c}^{t_{end}} \gamma(t)dt ~.
\label{df}\end{equation}
The end time $t_{end}$ is the time, when the mean free path,
$t_T$ equals $t_c$, the size of the perturbation.
We define $z_{dec}$ as the redshift when the
photons and baryons decouple due to free streaming. This is about the
time when the mean free path has grown up to the size of the horizon:
 $t_T(z_{dec}) \approx t(z_{dec})$. To obtain exponential
damping ($t_c < t_{end}$), we thus
need $z_c>z_{dec}$.     In this case damping is effective until
\[ 1+z_{end} = \sqrt{(1+z_{dec})(1+z_c)}~,  \]
 and we obtain
\begin{equation} f \approx 2\left({1+z_{dec}\over
1+z_c}\right)^{3/2}\left[\left(
      {1+z_c\over 1+z_{dec}}\right)^{9/4}-1\right]
      ~,  \label{damp}  \end{equation}
where we have set $k^2/6 = (2\pi/t_c)^2$.
In terms of angles this yields
\[ f(\theta) \approx 2(\theta/\theta_d)^3[(\theta_d/\theta)^{9/2} -1] ~,\]
where $\theta_d = 1/\sqrt{1+z_{dec}}\approx 5.7^0$.

If the plasma ionizes at a redshift $z_i$,  $z_{dec}<z_i<z_c$, after
the texture has already collapsed, damping is only effective after
$z_i$ if
$z_{end}< z_i$ and instead of formula (\ref{damp}) we then obtain
\begin{equation} f  \approx 2\left({1+z_{dec}\over 1+z_c}\right)^{3/2}
	\left[\left(
    {1+z_c\over 1+z_{dec}}\right)^{9/4}-\left(
    {1+z_c\over 1+z_i}\right)^{9/4}\right]~ . \label{damp2} \end{equation}
This damping factor can again be converted in an angular damping
factor in the above, obvious way.
The CBR signal of  textures is thus exponentially damped only if

$z_i>z_{dec} \approx 107(0.05/h_{50}\Omega_b)^{2/3}$  \\
and in this case only textures with
$ 1+z_{dec}< 1+z_c< (1+z_i)^2/(1+z_{dec})$ are  affected.
For exponential damping after reionization at  $z_i = 50$  extreme
values of the Hubble Constant and baryon densities far above that
predicted from big bang nucleosynthesis and light element
abundances are required.

In this  approximation we have neglected the the amount of damping
which still may occur after $z_{end}$.

\newpage

{\LARGE Figure Captions} \vspace{1cm}\\
{\bf Fig. 1:\ } The hot spot---cold spot signal of a spherically
symmetric collapsing texture in units of $\epsilon \sim 2.8\times 10^{-4}$.
  The horizontal variable $\tau = t-r\cos\theta$ denotes the 'impact time'
  (the time of closest encounter with the center of the texture)
  of a photon arriving at a distance $r$ from the texture at time $t$
  traveling with an angle $\theta$ with respect to the radial direction.
  The hot spot--cold spot is shown for photons with fixed impact parameter
  $b=r\sin\theta\approx 0.1t_c$ ($t_c$ is the time of texture collapse).

  A texture in a recombined expanding universe at $t=t_c$, line
(1) and    $t=1.5t_c$, line (2) is compared with the flat space
result (dashed curve).
  The second peak appearing at $t=1.5t_c$ is due to the dark
matter potential.
\vspace{1cm}\\
{\bf Fig. 2}\hspace{.3cm}
The CMB perturbation in units of $\epsilon \sim 2.8\times 10^{-4}$ as a
function of the impact parameter $b$ for fixed $\tau \sim 0.5t_c = 10$.
The signal disappears at an impact parameter $b\sim (1-1.5)t_c  $
($t_c = 20$ in the units chosen).
\vspace{1cm}\\
{\bf Fig. 3}\hspace{.3cm}
A comparison of $\Delta T/T$ from a compensated (solid line) and a not
compensated (dashed line) texture is shown. The not compensated signal
is acausal: a photon which had a distance $\approx 0.5t_c$ from the
texture at the time of the big bang ($t=0$) and aways was moving away from
it ($\tau\approx t_c/2$), i.e. never was in causal contact
with the
texture still  experiences about half the maximum energy shift. These
acausalities are removed in the compensated case.
\vspace{1cm}\\
{\bf Fig. 4}\hspace{.3cm}
A simulated COBE map as it might look in a scenario with texture + CDM.
The color scheme goes from $-4\times 10^{-4}$ (dark blue)  to
$4\times 10^{-4}$ (deep red). The zero point is not the middle of
the color palette but at the transition green/yellow. Monopole and
dipole contributions are subtracted in this map.
\vspace{1cm}\\
{\bf Fig. 5}\hspace{.3cm}
 The distribution of quadrupole values of 100 realizations of the
 microwave sky is shown in linear (a) and logarithmic (b) scale. The
 distribution is very broad with an average of $1.5\times 10^{-5}$
and a standard deviation of  $1.2\times 10^{-5}$.
\vspace{1cm}\\
{\bf Fig. 6}\hspace{.3cm}
The statistical distribution of the microwave anisotropies for one
realization of the microwave sky. The number of
pixels with a given $\Delta T/T$ are counted. The distribution is
slightly
non Gaussian, negatively skewed (skewness $= -3$, curtosis $=-1.2$).
\vspace{1cm}\\
{\bf Fig. 7}\hspace{.3cm}
The auto correlation function is presented in units $(\mu K)^2$ (for a bias
factor of 2, $\epsilon_0=1$, see text). In addition to the average values over
100 realizations (solid line), we indicate the $1\sigma$ upper and lower
limits (dashed lines).
\vspace{1cm}\\
{\bf Fig. 8}\hspace{.3cm}
The amplitudes of the spherical harmonics for $l=2$ to $64$, averaged
over 100 simulations are shown (fat solid line) and the $1\sigma$ upper and
lower limits are indicated (thin solid lines). For comparicon we have also
plotted the analytical fits for power law spectra with $n=1.5$
(short dashed line) and $n=1$ (long dashed line).
\vspace{1cm}\\
{\bf Fig. 9}\hspace{.3cm}
The spherical ring of positive--negative fluctuation as seen in the
microwave sky from a texture collapsing at redshift $z\approx 100$
a distance $r\approx t_0-t_c$ away from the observer.

\end{document}